\documentclass[
aps,
nofootinbib,
superscriptaddress,
tightenlines,
notitlepage,
twocolumn,
showpacs,
floatfix,
]{revtex4-1}
\usepackage{amssymb,amsmath,bm,tensor,braket}
\usepackage[colorlinks]{hyperref}

\usepackage{graphicx}
\usepackage{dcolumn}
\usepackage{bm}
\usepackage{xcolor}

\newcommand{\PlanckMass}{M_{\rm Pl}}

\newcommand{\Tabriz}{\affiliation{Faculty of Physics, University of Tabriz,
Tabriz 51666-16471, Iran}}
\newcommand{\KIAA}{\affiliation{Kavli Institute for Astronomy and
Astrophysics, Peking University, Beijing 100871, China}}
\newcommand{\NAOC}{\affiliation{National Astronomical Observatories,
Chinese Academy of Sciences, Beijing 100012, China}}

\begin{document}

\preprint{APS/123-QED}

\title{On the Cosmology of Dirac-Born-Infeld dRGT massive gravity}

\author{Sobhan Kazempour}\email{s.kazempour@tabrizu.ac.ir}\Tabriz
\author{Amin Rezaei Akbarieh}\email{am.rezaei@tabrizu.ac.ir}\Tabriz
\author{Hossein Motavalli}\email{Motavalli@tabrizu.ac.ir}\Tabriz
\author{Lijing Shao}\email{lshao@pku.edu.cn}\KIAA\NAOC

\date{\today}

\begin{abstract}
We introduce the cosmological analysis of the Dirac-Born-Infeld dRGT massive gravity theory which is a new extension of de Rham-Gabadadze-Tolley (dRGT) massive gravity. In this theory, we consider the Dirac-Born-Infeld (DBI) scalar field which is coupled to the graviton field. Moreover, we perform the cosmological background equations, and we demonstrate the self-accelerating background solutions. We show that the theory consists of self-accelerating solutions with an effective cosmological constant. In the following, we exhibit tensor perturbations analyses and achieve the dispersion relation of gravitational waves. We analyze the propagation of gravitational perturbation in the Friedmann-Lema\^itre-Robertson-Walker cosmology in the DBI dRGT massive gravity. Finally, we present the vector and scalar perturbations to show the stability conditions of the theory.
\end{abstract}

\maketitle


\section{\label{sec:intro}Introduction}

According to a number of observational evidence such as baryon acoustic oscillations \cite{Beutler:2011hx,SDSS:2009ocz}, CMB \cite{Planck:2015fie,WMAP:2003elm}, and supernovas I-a \cite{Phillips:1993ng,SupernovaSearchTeam:1998fmf}, it is strongly accepted that the Universe is in the accelerated expansion phase. To explain the origin of the accelerated expansion of the Universe and the cosmological constant problem, there are some different approaches \cite{Copeland:2006wr,Sotiriou:2008rp,Clifton:2011jh,Nojiri:2006ri,Bamba:2012cp}. It is clear that general relativity is a unique theory of a massless Lorentz-invariant spin-2 particle \cite{Weinberg:1965rz}. However, this theory can not explain the late-time acceleration of the Universe \cite{Weinberg:1988cp,Peebles:2002gy}. One approach to solving the problems in cosmology is a massive gravity theory in which gravity is propagated by a spin-2 nonzero graviton mass \cite{deRham:2010ik,deRham:2010kj,Hinterbichler:2011tt,deRham:2014zqa,Hassan:2011hr,Hassan:2011zd}.

In 1939, Fierz and Pauli introduced the first massive spin-2 field theory. They presented the unique Lorentz-invariant linear theory without ghosts in a flat spacetime \cite{Fierz:1939ix}.  
But, there was a discontinuity (van Dam-Veltman-Zakharov i.e., vDVZ discontinuity) which the theory does not reduce to general relativity in the limit of $m_{g} \rightarrow 0$ \cite{vanDam:1970vg,Zakharov:1970cc}. 
While Vainshtein solved this problem by considering the nonlinear Fierz-Pauli action instead of linear \cite{Vainshtein:1972sx}, Boulware and Deser claimed that the nonlinear Fierz-Pauli action has a ghost which is called the Boulware-Deser ghost \cite{Boulware:1972yco}.
Also, Arkani-Hamed et al. and Creminelli et al. confirmed this issue which the
nonlinear massive gravity is an unstable theory \cite{Arkani-Hamed:2002bjr,Creminelli:2005qk}.
However, de Rham, Gabadadze, and Tolley (dRGT) demonstrated the fully nonlinear
massive gravity without the Boulware-Deser ghost in 2010 \cite{deRham:2010ik,deRham:2010kj}.
They constructed a theory with nonlinear interactions which can show the massive spin-2 field in a flat spacetime.

As the dRGT massive gravity theory is only valid for an open FLRW solution, and there are not any stable solutions for homogeneous and isotropic Universe \cite{DeFelice:2012mx}, alternative theories have been proposed \cite{DAmico:2011eto,Gumrukcuoglu:2012aa,DeFelice:2013awa,Huang:2012pe,Hassan:2011zd,Hinterbichler:2012cn}.
In addition, in the dRGT massive gravity theory, there are a strong coupling problem and a nonlinear ghost instability, which is why the scalar and vector perturbations vanish \cite{Gumrukcuoglu:2011zh}. It is noticeable that considering the extra degree of freedom such as an extra scalar field is one of the great approaches. This way, the quasi-dilaton massive gravity theory is successful to explain the accelerated expansion of the Universe \cite{DAmico:2012hia}. However, because of the perturbations instability in the quasi-dilaton massive gravity theory, extensions of this theory have been presented \cite{DeFelice:2013tsa,Mukohyama:2014rca,Akbarieh:2021vhv,Aslmarand:2021qwn}.
In this paper, we consider an extra scalar field whose kinetic term has a Dirac-Born-Infeld (DBI) form. Actually, we introduce the new extension of the dRGT massive gravity theory. Also, we will show the perturbations analyses for this new extension in order to show this new extension is free of instability.

As there is an attractor solution in the scalar fields with inverse power-law potentials, there has been a tendency towards this kind of theory \cite{Ratra:1987rm}. In other words, the significance of this issue lies in the fact that the present-day behavior of the Universe is insensitive to the initial conditions \cite{Martin:2008xw}.
The Dirac-Born-Infeld model has something to do with inflation and string theory. Recently, there has been a trend toward finding the connection between string theory and inflation.  It should be mentioned that the main ideas are that according to the concept of brans, inflation interprets as the distance between two branes that move in the extra dimensions along a warped throat \cite{Martin:2008xw,Kachru:2003sx,Silverstein:2003hf,Alishahiha:2004eh,Lorenz:2007ze}.

As several attempts have been done for introducing the new extensions of massive gravity theories \cite{Gannouji:2013rwa,Mukohyama:2014rca,Gumrukcuoglu:2013nza,Gabadadze:2014kaa,Motohashi:2014una,Kahniashvili:2014wua,Gumrukcuoglu:2016hic,Gumrukcuoglu:2020utx,Akbarieh:2021vhv,Aslmarand:2021qwn}, we would like to introduce the new extension of dRGT massive gravity theory which can explain the accelerated expansion of the Universe in the FLRW cosmology, and we also try to show the conditions of stability of the system using the perturbations analyses. In the vector and scalar perturbations, we will exhibit the stability conditions of the system. Moreover, in the tensor perturbation, the dispersion relation of gravitational waves which can show the mass of graviton, will be obtained.

It should be noted that in order to impose the constraints on the modified gravity theories, detecting and analyzing gravitational waves is really essential. Using the mass of graviton, the speed of gravitational wave propagation can be determined. So, by comparing the speed of gravitational waves and their electromagnetic waves, we can find the constraints on the parameters of the theories. This issue shows the significance of the investigation of the dispersion relation of gravitational waves.

The outline of this paper is as follows. In Sec \ref{sec:1}, the Dirac-Born-Infeld dRGT massive gravity theory are presented. Moreover, we demonstrate the background equations of motion and self-accelerating solutions. In Sec \ref{sec:2}, we present perturbation analysis to determine the dispersion relations of gravitational waves in this theory. In addition, we elaborate on the stability condition of the system in the vector and scalar perturbations. In Sec \ref{sec:3}, we present the conclusion and discussion.

In this paper, we use $\PlanckMass^{2}\equiv 8\pi G =1$ where $G$ is Newton's gravitational constant.
It will be considered natural units ($c = \hslash = 1$).

\section{Dirac-Born-Infeld dRGT Massive Gravity}\label{sec:1}

In this section, we begin by introducing a new extension of the dRGT massive gravity theory. In this new extension, we consider the Dirac-Born-Infeld (DBI) scalar field which is the response to the dark energy scalar field.
We review the dRGT massive gravity theory which is extended by Dirac-Born-Infeld terms. Meanwhile, we perform the evolution of a cosmological background.

The action includes the Ricci scalar $R$, scalar field $\sigma$, the tension $T(\sigma)$, the massive gravity term $U({\cal K})$, a dynamical metric $g_{\mu\nu}$ and it's determinant $\sqrt{-g}$. The action is given by\\
\begin{eqnarray}\label{Action}
S=\frac{1}{2}\int d^{4} x \Bigg\{\sqrt{-g} \bigg[R + T(\sigma)\bigg(1-\sqrt{1-\frac{g^{\mu\nu}\partial_{\mu}\sigma\partial_{\nu}\sigma}{T(\sigma)}}\bigg)\nonumber\\+2{m}_{g}^{2}U(\mathcal K)\bigg]\Bigg\}.\nonumber\\
\end{eqnarray}
It should be pointed out that the mass of graviton $m_{g}$, originates from the potential $U(\mathcal{K})$ which is
\begin{eqnarray}\label{Upotential2}
U(\mathcal{K})&&=U_{2}+\alpha_{3}U_{3}+\alpha_{4}U_{4},
\nonumber\\
 U_{2}&&=\frac{1}{2}\big([\mathcal{K}]^{2}-[\mathcal{K}^{2}]\big),
 \nonumber\\
 U_{3}&&=\frac{1}{6}\big([\mathcal{K}]^{3}-3[\mathcal{K}][\mathcal{K}^{2}]+2[\mathcal{K}^{3}]\big),
 \nonumber\\
 U_{4}&&=\frac{1}{24}\big([\mathcal{K}]^{4}-6[\mathcal{K}]^{2}[\mathcal{K}^{2}]+8[\mathcal{K}][\mathcal{K}^{3}]+3[\mathcal{K}^{2}]^2\nonumber\\ && -6[\mathcal{K}^{4}]\big),
\end{eqnarray}
where $\alpha_3$ and $\alpha_4$ are dimensionless free parameters of the
theory.
Here the "$[\cdot]$'' is the trace of the tensor
inside brackets. Moreover, it is necessary to note that the building block tensor
$\mathcal{K}$ is defined as
\begin{equation}\label{K}
\mathcal{K}^{\mu}_{\nu} = \delta^{\mu}_{\nu} -
e^{\sigma}\sqrt{g^{\mu\alpha}f_{\alpha\nu}},
\end{equation}
where $ f_{\alpha\nu}$ is the fiducial metric, which is defined through
\begin{equation}\label{7}
f_{\alpha\nu}=\partial_{\alpha}\phi^{c}\partial_{\nu}\phi^{d}\eta_{cd}.
\end{equation}
Here $g^{\mu\nu} $ is the physical metric, $\eta_{cd}$ is the Minkowski
metric with $c,d= 0,1,2,3$ and $\phi^{c}$ are the Stueckelberg fields which
are introduced to restore general covariance. In addition, the theory is invariant under a global dilation transformation, $\sigma\rightarrow\sigma+\sigma_{0}$.

The FLRW Universe is considered because of our cosmological motivation.
As a result, we represent the dynamical and fiducial metrics,

\begin{align}
\label{DMetric}
g_{\mu\nu}&={\rm diag} \left[-N^{2},a^2,a^2,a^2 \right], \\
\label{FMetric} 
f_{\mu\nu}&={\rm diag} \left[-\dot{f}(t)^{2},1,1,1 \right].
\end{align}
It should be paid attention that $N$ is the lapse function of the
dynamical metric, and it is similar to a gauge function. Furthermore, the $a$ is the scale factor, and the $\dot{a}$ is the derivative with respect to time. Also, the lapse function has something to do with the coordinate-time $dt$ and the proper-time $d\tau$ via $d\tau=Ndt$
\cite{Scheel:1994yr,Christodoulakis:2013xha}. It is worth noting that the function $f(t)$ is the
Stueckelberg scalar function whereas $\phi^{0}=f(t)$ and
$\frac{\partial\phi^{0}}{\partial t}=\dot{f}(t)$ \cite{Arkani-Hamed:2002bjr}.

So, the point-like Lagrangian of the Dirac-Born-Infeld dRGT massive gravity in FLRW
cosmology is
\begin{eqnarray}
\mathcal{L}=\frac{-3 a\dot{a}^{2}}{N}-\frac{a^{3}N}{2}T(\sigma)\bigg[\sqrt{1-\frac{\dot{\sigma}^{2}}{T(\sigma) N^{2}}}-1\bigg]\nonumber\\+m_{g}^{2}a^{3}(Y-1)\Bigg\lbrace N\bigg[3(Y-2)-(Y-4)(Y-1)\alpha_{3}\nonumber\\-(Y-1)^{2}\alpha_{4}\bigg]+\dot{f}(t) a Y(Y-1)\bigg[3-3(Y-1)\alpha_{3}\nonumber\\+(Y-1)^{2}\alpha_{4}\bigg]\Bigg\rbrace,\nonumber\\
\end{eqnarray}
where
\begin{equation}\label{XX}
Y\equiv\frac{e^{\sigma}}{a}.
\end{equation}

To simplify expressions later, we define
\begin{equation}
H\equiv\frac{\dot{a}}{Na}.
\end{equation}

\subsection{Background Equations of Motion}\label{subsec4}

We obtain the cosmological background equations for a FLRW background.
This way, by varying with respect to $f$, and paying attention to the unitary gauge, i.e. $f(t)=t$,
a constraint equation is given by:
\begin{eqnarray}\label{Cons}
\frac{\delta \mathcal{L}}{\delta f}= m_{g}^{2} \frac{d}{dt}\bigg[  a^{4}Y(Y-1)
 \times\bigg(3-3(Y-1)\alpha_{3}\nonumber\\+(Y-1)^{2}\alpha_{4}\bigg)\Bigg]=0. \nonumber\\
\end{eqnarray}
On the classical level, the unphysical fields should be eliminated from the Lagrangian using gauge transformations \cite{Grosse-Knetter:1992tbp}.

In this step, we calculate the equation of motion which is related to the lapse function $N$
\begin{widetext}
\begin{eqnarray}\label{EqN}
\frac{1}{a^{3}}\frac{\delta \mathcal{L}}{\delta N}=3H^{2} + \frac{T(\sigma)}{2\bigg( T(\sigma) N^{2} + \big(H N+\frac{\dot{Y}}{Y}\big)^{2} \bigg)}\bigg[ \big(H N+\frac{\dot{Y}}{Y}\big)^{2} - T(\sigma) N^{2} \bigg( \sqrt{1+ \frac{\big(H N+\frac{\dot{Y}}{Y}\big)^{2}}{T(\sigma)N^{2}} } - 1 \bigg) \bigg] \nonumber\\ -m_{g}^{2}(Y-1)\bigg[-3(Y-2)+(Y-4)(Y-1)\alpha_{3}+(Y-1)^{2}\alpha_{4}\bigg]=0.\nonumber\\&&
\end{eqnarray}
\end{widetext}
By varying with respect to $\sigma$, we have
\begin{widetext}
\begin{eqnarray}\label{EqSig}
\frac{1}{a^{3}N}\frac{\delta S}{\delta \sigma}= \frac{1}{4 \big( H^{2} + T(\sigma)N^{2} \big)^{2}}\Bigg\lbrace  -2 T(\sigma)^{2}N^{4} T^{'}(\sigma) \bigg( \sqrt{1 + \frac{H^{2}}{T(\sigma) N^{2}}} - 1 \bigg)  + 2 H^{4} \bigg( 3 T(\sigma)N\sqrt{1+\frac{H^{2}}{T(\sigma)N^{2}}} + T^{'}(\sigma) \bigg) \nonumber\\ + H^{2}T(\sigma) N^{2}\bigg( 3\sqrt{1 + \frac{H^{2}}{T(\sigma)N^{2}}}\big( 2 T(\sigma)N - T^{'}(\sigma) \big) + 4 T^{'}(\sigma) \bigg) \Bigg\rbrace + m_{g}^{2}Y\Bigg\lbrace -(3+r)\big(3+3\alpha_{3}+\alpha_{4}\big) 
 \nonumber\\  +6(r+1)Y\big(\alpha_{4}+2\alpha_{3}+1\big)-3(3r+1)(\alpha_{4}+\alpha_{3})Y^{2}+4r\alpha_{4}Y^{3}\Bigg\rbrace =0,\nonumber\\
\end{eqnarray}
\end{widetext}
where
\begin{eqnarray}
r\equiv\frac{a}{N}.
\end{eqnarray}
Notice that the following equations can be obtained using Eq. (\ref{XX}),
\begin{equation}
\frac{\dot{\sigma}}{N}= H+\frac{\dot{Y}}{NY}, \qquad \ddot{\sigma}=\frac{d}{dt}\Big(NH+\frac{\dot{Y}}{Y}\Big),
\end{equation}
According to the time reparametrization invariance introduced by the Stueckelberg field $f$, 
there is a Bianchi identity that guarantees that the equation of motion related to the scale factor $a$ is redundant. The Bianchi identity is,
\begin{eqnarray}
\frac{\delta S}{\delta \sigma}\dot{\sigma}+\frac{\delta S}{\delta f}\dot{f}-N\frac{d}{dt}\frac{\delta S}{\delta N}+\dot{a}\frac{\delta S}{\delta a}=0.
\end{eqnarray}
Therefore, this equation can be ignored.

It should be indicated that in the spacial states, all of
the cosmological background equations reached
to those in Refs. \cite{DAmico:2012hia,Gumrukcuoglu:2013nza}.

\subsection{Self-Accelerating Background Solutions}\label{subsec5}

In this subsection, we indicate the analyses of self-accelerating solutions to explain the accelerated expansion of the Universe.
We now start the discussion of the solutions by the Stueckelberg constraint in Eq. (\ref{Cons}).
By integrating that equation we have,
\begin{eqnarray}\label{Self}
Y(Y-1)\bigg[3-3(Y-1)\alpha_{3}+(Y-1)^{2}\alpha_{4}\bigg] \propto a^{-4} .
\end{eqnarray}

It is necessary to mention that the constant solutions of $Y$ lead to
the effective energy density and behave similarly to a cosmological constant.
In other words, the massive gravity term affects the expansion like a cosmological constant.
As we would like to consider an expanding universe, and the right-hand side of the equation (\ref{Self}) will decrease after a long enough time, $Y$ saturates to a constant value $Y_{\rm SA}$, which is a root of the left-hand side of Eq.~(\ref{Self}).

According to \cite{DAmico:2012hia}, there are four constant solutions for $Y$.
For avoiding to encounter strong coupling, we leave this solution \cite{DAmico:2012hia}. 
Thus, we are left with,
\begin{equation}
\hspace{-0.4cm}
(Y-1)\big[3-3(Y-1)\alpha_{3}+(Y-1)^{2}\alpha_{4}\big]\bigg|_{Y=Y_{\rm SA}}=0.
\end{equation}
Furthermore, it is clear that the other solution is $Y=1$. However, this solution leads to inconsistency and vanishing cosmological constant, and this is the reason why it is unacceptable \cite{DAmico:2012hia}. 

So, the two remaining solutions of Eq. (\ref{Self}) are
\begin{equation}\label{XSA}
Y_{\rm SA}^{\pm}=\frac{3\alpha_{3}+2\alpha_{4}\pm\sqrt{9\alpha_{3}^{2}-12\alpha_{4}}}{2\alpha_{4}}.
\end{equation}
Here, we obtain the modified Friedmann equation. We represent the Friedmann equation (\ref{EqN}) in a different form,
\begin{eqnarray}\label{EqFr1}
3 H^{2} + \frac{\tilde{T}}{2 \big( \tilde{T} N^{2} +H^{2}N^{2} \big)}\bigg[ H^{2}N^{2} - \tilde{T} N^{2}\big( \sqrt{1 + \frac{H^{2}}{\tilde{T}}} \nonumber\\ - 1 \big) \bigg]  = \Lambda_{\rm SA}^{\pm}, \nonumber\\
\end{eqnarray}
where $\tilde{T}$ is a saturate of $T(\sigma)$.
We solve the Eq. (\ref{EqFr1}) to calculate the $H^{2}$, so we have
\begin{widetext}
\begin{eqnarray}
H^{2}=\frac{1}{18}\Bigg\lbrace4 (\Lambda_{\rm SA}^{\pm}-2 \tilde{T}) + \big( 2 \Lambda_{\rm SA}^{\pm} + 5 \tilde{T}\big)^{2}\bigg( 9\sqrt{\tilde{T}^{3}\big( -16 \Lambda_{\rm SA}^{\pm 3} -120 \Lambda_{\rm SA}^{\pm 2}\tilde{T} -300 \Lambda_{\rm SA}^{\pm} \tilde{T}^{2}- 169 \tilde{T}^{3} \big)} -2 \big( 4 \Lambda_{\rm SA}^{\pm 3} \nonumber\\ + 30 \Lambda_{\rm SA}^{\pm 2} \tilde{T}+ 75 \Lambda_{\rm SA}^{\pm} \tilde{T}^{2} + 22\tilde{T}^{3} \big) \bigg)^{-\frac{1}{3}} + \bigg( -2 \big(  4 \Lambda_{\rm SA}^{\pm 3} + 30 \Lambda_{\rm SA}^{\pm 2} \tilde{T}+ 75 \Lambda_{\rm SA}^{\pm} \tilde{T}^{2} + 22\tilde{T}^{3} \big) \nonumber\\ +  9\sqrt{\tilde{T}^{3}\big( -16 \Lambda_{\rm SA}^{\pm 3} -120 \Lambda_{\rm SA}^{\pm 2}\tilde{T} -300 \Lambda_{\rm SA}^{\pm} \tilde{T}^{2}- 169 \tilde{T}^{3} \big)}  \bigg)^{\frac{1}{3}} \Bigg\rbrace.
\end{eqnarray}
\end{widetext}
Note that the effective cosmological constant from the
mass term is
\begin{eqnarray}
\Lambda_{\rm SA}^{\pm}\equiv m_{g}^{2}(Y_{\rm SA}^{\pm}-1)\bigg[ && -3Y_{\rm SA}^{\pm} +6+(Y_{\rm SA}^{\pm}-4)(Y_{\rm SA}^{\pm}-1)\alpha_{3}\nonumber\\
&&+(Y_{\rm SA}^{\pm}-1)^{2}\alpha_{4}\bigg].
\end{eqnarray}
Using Eq. (\ref{XSA}), the above equation can be written as 
\begin{eqnarray}
\Lambda_{\rm SA}^{\pm}=\frac{3m^{2}_{g}}{2\alpha^{3}_{4}}\bigg[9\alpha^{4}_{3}\pm 3\alpha^{3}_{3}\sqrt{9\alpha^{2}_{3}-12\alpha_{4}}-18\alpha^{2}_{3}\alpha_{4}\nonumber\\\mp 4\alpha_{3}\alpha_{4}\sqrt{9\alpha^{2}_{3}-12\alpha_{4}}+6\alpha^{2}_{4}\bigg].
\end{eqnarray}
Therefore, from Eq.~(\ref{EqSig}) we have,
\begin{widetext}
\begin{eqnarray}\label{rS}
r_{\rm SA}= \frac{1}{2 \big( H^{2} + \tilde{T} \big)} \Bigg\lbrace 2 \tilde{T} + H^{2} \bigg( 2 - \frac{\tilde{T}N\sqrt{1+\frac{H^{2}}{\tilde{T}N^{2}}}}{m_{g}^{2}Y_{\rm SA}^{\pm \rm 2}\big( \alpha_{3}Y_{\rm SA}^{\pm} - \alpha_{3} -2 \big)} \bigg) \Bigg\rbrace . \nonumber\\
\end{eqnarray}
\end{widetext}
The above equation interprets the self-accelerating universe without any strong coupling.
Thus, we have shown that this theory consists of self-accelerating
solutions with an effective cosmological constant.

\section{PERTURBATIONS ANALYSIS}\label{sec:2}

At this stage, we would like to demonstrate the perturbations analyses. These analyses are crucial to indicate the stability of solutions. 
To reach this goal, we focus on quadratic perturbations. 
The physical metric $g_{\mu\nu}$ can be expanded in terms of small fluctuations $\delta g_{\mu\nu}$ around a background solution $g_{\mu\nu}^{(0)}$.
\begin{equation}
g_{\mu\nu}=g_{\mu\nu}^{(0)}+\delta g_{\mu\nu}.
\end{equation}
Note that the metric perturbations can be divided into three parts,
namely scalar, vector, and tensor perturbations. Therefore, we have
\begin{eqnarray}
\delta g_{00}=&&-2N^{2} \Phi, \nonumber\\
\delta g_{0i}=&&Na(B_{i}+\partial_{i}B), \nonumber\\
\delta g_{ij}=&&a^{2}\bigg[h_{ij}+\frac{1}{2}(\partial_{i}E_{j}+\partial_{j}E_{i})+2\delta_{ij}\Psi \nonumber\\
&& +\big(\partial_{i}\partial_{j} -\frac{1}{3}\delta_{ij}\partial_{l}\partial^{l}\big)E\bigg],
\end{eqnarray}
All perturbations agree with the spatial rotation transformations, and they are functions of time and space. Also, there are these conditions $\delta^{ij}h_{ij}=\partial^{i}h_{ij}=\partial^{i}E_{i}=\partial^{i}B_{i}=0$ for scalar, vector, and tensor perturbations which means that the tensor perturbations are transverse and traceless.

The the scalar field $\sigma$ would be perturbed as follows
\begin{equation}
\sigma =\sigma^{(0)} + \delta\sigma.
\end{equation}
In the following procedures, all terms should be kept in quadratic order in $\delta
g_{\mu\nu}$. It is crucial to mention that we have not had any problems with the form of gauge-invariant combinations because we have indicated all analyses in the unitary gauge. 
In addition, the expanded actions can be written in the Fourier domain with plane waves, i.e., $\vec{\nabla}^{2}\rightarrow -k^{2}$, $d^{3}x\rightarrow
d^{3}k$. Note that the spatial indices on perturbations are raised and lowered by $\delta^{ij}$ and $\delta_{ij}$.

\subsection{Tensor}\label{subTen}

The significance of tensor perturbation analysis lies in the fact that it is the only source of gravitational waves in general relativity. As the dispersion relation of gravitational waves is different in modified gravity models, the speed of propagation of the gravitational wave is different from the speed of light in standard general relativity.
In this subsection, we obtain the dispersion relation of gravitational waves in the Dirac-Born-Infeld dRGT massive gravity theory.
We consider tensor perturbations around the background,
\begin{equation}
\delta g_{ij}=a^{2}h_{ij},
\end{equation}
where
\begin{equation}
\partial^{i}h_{ij}=0 \quad {\rm and} \quad g^{ij}h_{ij}=0.
\end{equation}
Note that we calculate separately the tensor perturbed actions in the second-order for any part of the main action. The gravity part of the perturbed action in quadratic order is
\begin{eqnarray}
S^{(2)}_{\rm gravity}=&&\frac{1}{8}\int d^{3}k \, dt \, a^{3}\bigg[\frac{\dot{h}_{ij}\dot{h}^{ij}}{N^{2}}\nonumber\\&&-\Big(\frac{k^{2}}{a^{2}}+4\frac{\dot{H}}{N}+6H^{2} \Big)h^{ij}h_{ij}\bigg].
\end{eqnarray}
Furthermore, we obtain the Dirac-Born-Infeld part of the perturbed action in quadratic order
\begin{eqnarray}
S^{(2)}_{\rm DBI}=\frac{1}{8}\int d^{3}k \, dt \, a^{3}N \bigg[T(\sigma)\bigg( \sqrt{1-\frac{\dot{\sigma}^{2}}{T(\sigma)N^{2}}}-1\bigg) \bigg] h^{ij}h_{ij}. \nonumber\\
\end{eqnarray}
We derive the massive gravity sector of the perturbed action
\begin{eqnarray}
S^{(2)}_{\rm massive}&&= \frac{1}{8}\int d^{3}k \, dt \, a^{3}N m_{g}^{2}\bigg[(\alpha_{3}+\alpha_{4})rY^{3}-(1\nonumber\\&&+2\alpha_{3}+\alpha_{4})(1+3r)Y^{2}+(3+3\alpha_{3}+\alpha_{4})\nonumber\\
&&(3+2r)Y-2(6+4\alpha_{3}+\alpha_{4})\bigg]h^{ij}h_{ij}.
\end{eqnarray}
Summing up the second-order pieces of the perturbed actions $S^{(2)}_{\rm gravity}$, $S^{(2)}_{\rm DBI}$, and $S^{(2)}_{\rm massive}$, we calculate the total action in the second-order for tensor perturbations
\begin{eqnarray}
S^{(2)}_{\rm total}=\frac{1}{8}\int d^{3}k \, dt \, a^{3}N\bigg\lbrace \frac{\dot{h}_{ij}\dot{h}^{ij}}{N^{2}}-\Big(\frac{k^{2}}{a^{2}}+M_{\rm GW}^{2}\Big)h^{ij}h_{ij}\bigg\rbrace . \nonumber\\
\end{eqnarray}
At this stage, using Eq. (\ref{XSA}), we calculate $\alpha_{3}$ and substitute it. As a result, the dispersion relation of gravitational waves is obtained as 
\begin{eqnarray}\label{eq:M2:GW1}
M^{2}_{\rm GW}=4\frac{\dot{H}}{N}+6 H^{2}- T(\sigma)\bigg( \sqrt{1-\frac{\dot{\sigma}^{2}}{T(\sigma)N^{2}}}-1\bigg) - \Delta ,\nonumber\\   
\end{eqnarray}
where
\begin{widetext}
\begin{eqnarray}
\Delta = \frac{m_{g}^{2}}{(Y_{\rm SA}^{\pm}-1)} \Bigg\lbrace Y_{\rm SA}^{\pm}\bigg[ 18 + 8 \alpha_{3} + Y_{\rm SA}^{\pm} \bigg( 2\alpha_{3}Y_{\rm SA}^{\pm} + Y_{\rm SA}^{\pm} - r_{\rm SA}\big( 3(\alpha_{3}+2) \nonumber\\+ Y_{\rm SA}^{\pm}(\alpha_{3}Y_{\rm SA}^{\pm} -4\alpha_{3}-3)\big) -8\alpha_{3}-10\bigg)\bigg] -2(\alpha_{3}+3) \Bigg\rbrace, \nonumber\\
\end{eqnarray}
\end{widetext}
where $Y_{\rm SA}^{\pm}$ can not be equal to 1, because the role of the massive gravity term vanishes.

It is noticeable that if the mass square of gravitational waves would be positive, the stability
of long-wavelength gravitational waves is guaranteed. However, if it would be negative, it is tachyonic. So, the instability can take the age of the Universe to develop if the mass of the tachyon would be the order of the Hubble scale.

In fact, we exhibit the modified dispersion relation of gravitational waves.
Actually, the propagation of gravitational perturbations in the FLRW cosmology in the
Dirac-Born-Infeld dRGT massive gravity is presented.

\subsection{Vector}

Here the vector perturbation analyses are shown for the Dirac-Born-Infeld dRGT massive gravity theory. This part's main goal is to show the conditions in which there would not be any instabilities.

We consider the vector perturbations,
\begin{eqnarray}\label{Bi}
B_{i}=\frac{a(r^{2}-1)k^{2}}{\bigg[ 2 k^{2}(r-1) + 4 T(\sigma)\big( \sqrt{\frac{H^{2}a^{2}}{T(\sigma)N^{2}}} - 1 \big) \bigg]}\frac{\dot{E}_{i}}{N}.
\end{eqnarray}
As the field $B_{i}$ is nondynamical, it can be entered into the action as an auxiliary field.
So, a single propagating vector is given
\begin{eqnarray}\label{AVc}
S_{\rm vector}^{(2)}=\frac{1}{8}\int d^{3}k \, dt \, a^{3}N 
\bigg(\frac{\beta}{N^{2}} |\dot{E}_{i}|^{2} -\frac{k^{2}}{2}M_{\rm GW}^{2}|E_{i}|^{2}\bigg),\nonumber\\
\end{eqnarray}
where
\begin{eqnarray}\label{Betta}
\beta =\frac{k^{2}}{2}\bigg(1+\frac{k^{2}(r^{2}-1)}{2 T(\sigma)( \sqrt{\frac{H^{2}a^{2}}{T(\sigma)N^{2}}} - 1) }\bigg)^{-1}.
\end{eqnarray}
It can be found that there are two cases, in the first one, if we have $\frac{(r^{2}-1)}{2T(\sigma)\big( \sqrt{\frac{H^{2}a^{2}}{T(\sigma)N^{2}}} - 1 \big)}\geq 0$, we do not need the critical momentum scale.
However, in the second one, if we have $\frac{(r^{2}-1)}{2T(\sigma)\big( \sqrt{\frac{H^{2}a^{2}}{T(\sigma)N^{2}}} - 1 \big)}<0$, we should have a critical momentum scale $k_{c}$ to avoid a ghost, which is 
\begin{eqnarray}\label{39}
k_{c}= \sqrt{\frac{2 T(\sigma)\big[ \sqrt{\frac{H^{2}a^{2}}{T(\sigma)N^{2}}} - 1 \big]}{(1 - r^{2} )}} \quad {\rm if} \quad \frac{(r^{2}-1)}{2T(\sigma)\big( \sqrt{\frac{H^{2}a^{2}}{T(\sigma)N^{2}}} - 1 \big)}<0, \nonumber\\
\end{eqnarray}
in other words, to have stability in the system we require the $k$ in Eq. (\ref{Betta}) to be smaller than a critical momentum scale $k_{c}$.

In the following, we consider the canonically normalized fields for determining the other instabilities in the vector modes as below
\begin{eqnarray}\label{Khi}
\zeta_{i}=\frac{\beta E_{i}}{2}.
\end{eqnarray}
We have substituted Eq. (\ref{Khi}) to the Eq. (\ref{AVc}),
\begin{eqnarray}
S=\frac{1}{2}\int d^{3}k \, dt \, a^{3}N \bigg(\frac{|\dot{\zeta_{i}}|^{2}}{N^{2}}-c_{V}^{2}|\zeta_{i}|^{2}\bigg).
\end{eqnarray}
The sound speed for vector modes is given by
\begin{eqnarray}\label{c_V}
c_{V}^{2}=M_{\rm GW}^{2}(1+u^{2})-\frac{H^{2}u^{2}(1+4u^{2})}{(1+u^{2})^{2}},
\end{eqnarray}
where the dimensionless quantity is considered as below
\begin{eqnarray}
u^{2}\equiv \frac{k^{2}(r^{2}-1)}{2 T(\sigma) \big( \sqrt{\frac{H^{2}a^{2}}{T(\sigma)N^{2}}} - 1 \big)}.
\end{eqnarray}
In fact, we evaluate the conditions which can cause instability in the system.
Notice that to avoid tachyonic instability which can be originated from the first part of Eq. (\ref{c_V}), if we have $M_{\rm GW}^{2}<0$ and $u^{2}>0$, and to avoid instability the below conditions must be there.
\begin{eqnarray}
&& k_{c}^{2} \lesssim \frac{2 T(\sigma) \big( \sqrt{\frac{H^{2}a^{2}}{T(\sigma)N^{2}}} - 1 \big)}{(r^{2} - 1)}, \nonumber\\ && {\rm if} \quad \frac{(r^{2}-1)}{2T(\sigma)\big( \sqrt{\frac{H^{2}a^{2}}{T(\sigma)N^{2}}} - 1 \big)}>0 \quad {\rm and} \quad M_{\rm GW}^{2}<0.
\end{eqnarray}
If we consider all physical momenta below a critical momentum scale $k_{c}$, thus, a growth rate of instability should be lower than the cosmological scale.

Furthermore, if we pay attention to the second part of Eq. (\ref{c_V}), two cases can be considered. In the first case, if we have $u^{2}>0$, there is not any instability faster than the Hubble expansion.
In the second case which is $u^{2}<0$, as we have the no-ghost condition Eq. (\ref{39}), to avoid instability we should have $|u^{2}|\lesssim \frac{k^{2}}{k_{c}^{2}}$. This way, there is not any instability in the second part of Eq. (\ref{c_V}).

In the end, in order to guarantee the stability for vector modes, we should consider $c_{V}^{2}>0$.
In addition, for avoiding the instability the mass square of the dispersion relation of gravitational waves should be positive $M_{\rm GW}^{2}>0$, as we pointed out in the Sec. \ref{subTen}.

\subsection{Scalar}

In this subsection, we exhibit the scalar perturbations analyses in the Dirac-Born-Infeld dRGT massive gravity theory to elaborate on the stability of the system. 

The process would be started by the action quadratic in scalar perturbations
\begin{eqnarray}
\delta g_{00}=&&-2N^{2} \Phi, \nonumber\\
\delta g_{0i}=&&N\,a\,\partial_{i}B, \nonumber\\
\delta g_{ij}=&&a^{2}\bigg[2\delta_{ij}\Psi +\big(\partial_{i}\partial_{j} -\frac{1}{3}\delta_{ij}\partial_{l}\partial^{l}\big)E\bigg],
\end{eqnarray}
\begin{equation}
\sigma =\sigma^{(0)}+\delta\sigma.
\end{equation}
According to the fact that the perturbation $\Phi$ and $B$ are free of time derivatives, they can be used as auxiliary fields using their equations of motion.
\begin{eqnarray}
B=\frac{(r^{2}-1)a}{3 T(\sigma)\bigg[ 1 - \sqrt{\frac{H^{2}a^{2}}{T(\sigma)N^{2}}} \bigg] }\Bigg\lbrace 6 H \Phi - \frac{1}{N}(k^{2}\dot{E}+6\dot{\Psi}) \nonumber\\ + \frac{3T(\sigma)}{H a^{2}}\bigg[ 1 - \sqrt{\frac{H^{2}a^{2}}{T(\sigma)N^{2}}} \bigg] \delta\sigma \Bigg\rbrace, \nonumber\\
\end{eqnarray}

\begin{widetext}
\begin{eqnarray}
\Phi =\frac{1}{ 3 H^{4} \bigg[ 4 k^{2}\big( r^{2} - 1 \big) + \frac{1}{H^{2}a^{2}} \big( 2 Q T(\sigma) - T(\sigma)^{2} \big) + 6 Q - 7 T(\sigma)  \bigg]} \Bigg\lbrace \frac{k^{4}}{a^{4}} \bigg( 2 k^{2} \big( r^{2} -1 \big) \nonumber\\ - \frac{3}{(r - 1)} \big( Q - T(\sigma) \big) \bigg) \bigg( T(\sigma) - Q \bigg)^{2} \big( E + 3 \big) \delta\sigma + \frac{3 T(\sigma) H^{2}}{(r-1)a^{2}} \bigg[ \bigg( 2 k^{2} (r-1) \nonumber\\ + 3 \big( Q - T(\sigma) \big) \bigg)  \bigg] \bigg( \frac{Q}{T(\sigma)} - 1 \bigg) \Psi + \frac{2 H^{3}k^{2}}{N}\bigg( k^{2}\dot{E} + 6 \dot{\Psi} \bigg) \big( r^{2} -1 \big) \nonumber\\ - \frac{3 H T(\sigma)}{N a^{2}} \bigg( \big( Q - T(\sigma) \big) \delta \dot{\sigma} - 6 H^{2}a^{2} \dot{\Psi} \bigg) \big( \frac{Q}{T(\sigma)} - 1 \big) \Bigg\rbrace, \nonumber\\
\end{eqnarray}
\end{widetext}
where
\begin{eqnarray}
Q \equiv \sqrt{\frac{T(\sigma)H^{2}a^{2}}{N^{2}}}.
\end{eqnarray}
It is worth pointing out that if we substitute these equations into the action, we obtain the action which contains three fields, $E$, $\Psi$, and $\delta\sigma$.
Meanwhile, we determine another nondynamical combination to remove the sixth degree of freedom, which is
\begin{eqnarray}
\tilde{\Psi}= \frac{1}{\sqrt{2}}(\Psi +\delta\sigma).
\end{eqnarray}
Moreover, an orthogonal combination can be defined,
\begin{eqnarray}
\tilde{\delta\sigma}=\frac{1}{\sqrt{2}k^{2}}(\Psi -\delta\sigma).
\end{eqnarray}
Note that using above field redefinitions, we write the action in terms of $\tilde{\Psi}$, $\tilde{\delta\sigma}$, and $E$, with no time derivatives on $\tilde{\Psi}$. Using the following equation the auxiliary of $\tilde{\Psi}$ can be eliminated,
\begin{widetext}
\begin{eqnarray}
\tilde{\Psi}= \frac{1}{3 r (r -1) \bigg[ \big( T(\sigma)^{2} - 2 Q T(\sigma) \big) +H^{2} a^{2} \bigg( 4 k^{2} +\big( 7 T(\sigma) -6 Q \big) \bigg) \bigg] } \Bigg\lbrace  \bigg[ - 3 k^{2} r \big( r - 1 \big) \bigg( 2 Q T(\sigma) \nonumber\\ - T(\sigma)^{2} \bigg) + 72 H^{4} a^{4} \bigg( 4 k^{2} \big( r^{2} -1 \big) +\big( 6 Q - 7 T(\sigma) \big) \bigg) - 3 H^{2} a^{2} \bigg( 4 k^{4} r \big( r - 1 \big) + k^{2} r \big( 6 Q  \nonumber\\ + 6 r Q - T(\sigma) ( 7 r - 5 ) \big) + 24 \big( T(\sigma)^{2} - 2 Q T(\sigma) \big) \bigg) \bigg] \tilde{\delta\sigma}- 2 \sqrt{2} H^{2} a^{2} k^{4} r \big( r - 1 \big) E \nonumber\\ + \bigg[ 18 H^{3} a^{4} \bigg( 4 k^{2} \big( r^{2} - 1 \big) + 6 Q - 7 T(\sigma)  \bigg) + 6 H a^{2} \bigg( -2 k^{2} r ( r - 1) \big( Q - T(\sigma) \big) \nonumber\\ - 3 T(\sigma)^{2}  + 6 Q T(\sigma)  \bigg) \bigg] \frac{\dot{\tilde{\delta\sigma}}}{N} + \sqrt{2}H a^{2}k^{2} r \big( r - 1 \big)  \bigg[ 6 H^{2} a^{2} + T(\sigma) - Q \bigg] \frac{\dot{E}}{N} \Bigg\rbrace. \nonumber\\
\end{eqnarray}
\end{widetext}
We substitute this solution in the action, and by considering the notation $A \equiv (\tilde{\delta\sigma}, E)$, the scalar action is given by,
\begin{eqnarray}
S = \frac{1}{2}\int d^{3}k \, dt \, a^{3}N\Bigg\lbrace \frac{\dot{A}^{\dagger}}{N}\mathcal{F}\frac{\dot{A}}{N}+\frac{\dot{A}^{\dagger}}{N}\mathcal{D}A \nonumber\\+ A^{\dagger}D^{T}\frac{\dot{A}}{N}-A^{T}\varpi^{2}A\Bigg \rbrace,
\end{eqnarray}
here, we should pay attention that $D$ is a real anti-symmetric $2\times 2$ matrix, and $\mathcal{F}$ and $\varpi^{2}$ are real symmetric $2 \times 2$ matrices.

This way, we demonstrate the components of the matrix $\mathcal{F}$ as below
\begin{widetext}
\begin{eqnarray}
\mathcal{F}_{11}= \frac{2 k^{4}\big( Q - 1 \big)}{H^{2}a^{2}} \Bigg[1 + \frac{9 H^{2}a^{2}}{k^{2}\big( r - 1 \big)^{2}} - \frac{\bigg( \big( Q  - T(\sigma) \big) (r-1) - 6 H^{2} a^{2} r \bigg)^{2}}{\bigg[ \big( T(\sigma)^{2} - 2 Q T(\sigma) \big) + H^{2}a^{2}\bigg( 4 k^{2} + \big( 7 T(\sigma) - 6 Q  \big) \bigg) \bigg] \big( r - 1 \big)^{2} } \Bigg],
\end{eqnarray}
\begin{eqnarray}
\mathcal{F}_{12}= \frac{\sqrt{2}k^{4}\big( Q - T(\sigma) \big) \bigg[ 2 k^{2}\big( r - 1 \big) -3 r \big( 6 H^{2}a^{2} + T(\sigma) - Q\big) \bigg] }{3 \big( r - 1 \big) \bigg[ \big( T(\sigma)^{2} -2 T(\sigma) Q \big) +H^{2}a^{2}\big( 4 k^{2} + 7 T(\sigma) -6 Q \big) \bigg] },
\end{eqnarray}
\begin{eqnarray}
\mathcal{F}_{22}=   \frac{k^{4} \big( Q - T(\sigma) \big) \Bigg[ 2 k^{2} - 18 H^{2}a^{2} + 3\big( Q - T(\sigma) \big) \Bigg]}{18 \big( T(\sigma)^{2} - 2 Q T(\sigma) \big) + 18 H^{2}a^{2}\bigg[ 4 k^{2} + 7 T(\sigma) - 6 Q \bigg] }.
\end{eqnarray}
\end{widetext}
For determining the sign of the eigenvalues, we elaborate the determinant of the kinetic matrix $\mathcal{F}$. Therefore, we have
\begin{eqnarray}\label{69}
{\rm det} \, \mathcal{F}\equiv &&\mathcal{F}_{11}\mathcal{F}_{22}-\mathcal{F}_{12}^{2}=\nonumber\\ && \frac{3 k^{6} \bigg( T(\sigma) - Q \bigg)^{2} }{(r-1)^{2}\bigg(  Q - T(\sigma) - \frac{4 k^{2}H^{2}a^{2}}{6 H^{2}a^{2}+ T(\sigma) - Q } \bigg) }.
\end{eqnarray}
To avoid appearing the ghosts in the scalar sector, the following condition should be taken into consideration,
\begin{widetext}
\begin{eqnarray}
k < \frac{1}{2 a H}\sqrt{\bigg( 6 H^{2}a^{2}+T(\sigma) - \sqrt{\frac{H^{2}a^{2}T(\sigma)}{N^{2}}} \bigg) \bigg( \sqrt{\frac{H^{2}a^{2}T(\sigma)}{N^{2}}} - T(\sigma) \bigg)}. \nonumber\\
\end{eqnarray}
\end{widetext}
Thus, if the determinant is positive, we do not have a ghost degree of freedom. In other words, it should be mentioned that the stability of the scalar sector could be guaranteed with the use of the determinant of the kinetic matrix. We have shown the conditions under which the system can maintain its stability.

\section{Conclusion}\label{sec:3}

In conclusion, we should like to point out that introducing the new extended dRGT massive gravity theory is so important due to this fact that can give us the opportunity to understand how the extended theory can behave around their cosmological backgrounds.
In this present work, we have presented the Dirac-Born-Infeld dRGT massive gravity which is constructed by considering coupled DBI scalar field to the graviton field.

Initially, we have obtained the details of the new action and total Lagrangian. We also performed the full set of equations of motion for a FLRW background. In order to explain the late-time acceleration of the Universe in the context of the Dirac-Born-Infeld dRGT massive gravity, we exhibited the self-accelerating background solutions. In fact, we have shown that the theory consists of self-accelerating solutions with an effective cosmological constant which has something to do with massive gravity term.

Finally, we have presented the cosmological perturbations analyses, which consist of tensor, vector, and scalar perturbations.
In tensor perturbation, we have demonstrated the tensor perturbation calculation to investigate the mass of graviton for the Dirac-Born-Infeld dRGT massive gravity theory. In other words, we have presented the modified dispersion relation of gravitational waves. Furthermore, we have indicated that to guarantee the stability of long-wavelength, the square of gravitational waves should be positive i.e., $M_{\rm GW}^{2}>0$. Presenting the propagation of gravitational perturbation in the FLRW cosmology in the Dirac-Born-Infeld dRGT massive gravity theory is really vital in the era of gravitational waves.
Empirically, the mass of graviton can be constrained by many means~\cite{deRham:2016nuf}, including the propagation of gravitational waves~\cite{Will:1997bb, LIGOScientific:2021sio}, the dynamics of Solar system~\cite{Will:2018gku}, the timing of binary pulsars~\cite{Finn:2001qi, Miao:2019nhf, Shao:2020fka}, and so on. These methods probe different aspects of massive gravity theories~\cite{deRham:2016nuf}.
At the end, we have evaluated the vector and tensor perturbations to determine the stability conditions of the system.

\section*{Acknowledgements}
We are really grateful to Nishant Agarwal for helpful notes and codes which are related to tensor perturbations.
Also, S.K and A.R.A would like to thank A.\ Emir Gumrukcuolu for his useful comments.
L.S was supported by the National Natural Science Foundation of China (Nos. 11991053, 11975027, 11721303), the National SKA Program of China (No. 2020SKA0120300), and the Max Planck Partner Group Program funded by the Max Planck Society.


\bibliography{apssamp}


\end{document}